\documentclass[12pt]{article}

\newtheorem{lemma}{Lemma}
\newtheorem{theorem}{Theorem}

\begin{document}

\title{ \bf \Large Orthonormal Polynomials on the Unit Circle and 
 Spatially Discrete Painlev\'e II Equation}

\author{ 
 \small  Chie Bing Wang\thanks{ The work is partially supported 
  by NSF Grant DMS-9802122. } \\
 \small Department of Mathematics  \\
 \small  and \\ 
 \small Institute of Theoretical Dynamics \\
 \small University of California, Davis, CA 95616 \\
 \small email address: wang@itd.ucdavis.edu }

\date{}
\maketitle

\begin{abstract}
We consider the  polynomials 
$\phi_n(z)= \kappa_n (z^n+ b_{n-1} z^{n-1}+ \cdots)$
orthonormal with respect to the weight 
$\exp(\sqrt{\lambda} (z+ 1/z)) dz/2 \pi i z$ 
on the unit
circle in the complex plane. The leading coefficient $\kappa_n$ is found to
satisfy a difference-differential (spatially discrete) 
 equation which is further proved
to approach a third order differential equation 
 by double scaling.
The third order differential equation  is equivalent to
the Painlev\'e II equation. 
The leading coefficient and second leading coefficient of
$\phi_n(z)$ can be expressed asymptotically in terms of 
the Painlev\'e II function.
\end{abstract}

\setcounter{equation}{0}
 \section{ Introduction } 

   In this paper, we discuss the orthonormal polynomials with respect
to the weight $\exp(\sqrt{\lambda} (z+ 1/z)) dz/2 \pi i z$ on the unit
circle in the complex plane \cite{szego},
\begin{equation}
 \oint \phi_n (z) \overline{\phi_m (z)} e^{\sqrt{\lambda} (z+{1\over z})} {dz \over 2 \pi i z}
= \delta_{nm}, \,\,\,\,\,\, m,n \ge 0 ,  \label{E:1.1}
\end{equation}   
where the integral is over the unit circle, and $\overline{\phi_m(z)}$ means the complex
conjugate of $\phi_m(z)$, $\lambda$ is a positive  parameter.
The polynomials $\phi_n (z) =\kappa_n  z^n + \cdots$ have explicit representation
in terms of the Toeplitz determinants 
$D_n (\lambda) = D_n \left( \exp (2 \sqrt{\lambda} \cos \theta) \right)$ \cite{szego}.
Here we are interested in analyzing the properties of the leading coefficient
$\kappa_n (\lambda)$.  As discussed in \cite{szego}, $\kappa_n^2$ can be expressed
in terms of the Toeplitz determinants,
\begin{equation}
 \kappa_n^2 = {D_{n-1} (\lambda) \over D_{n} (\lambda) }.  \label{E:1.1a}
\end{equation}
And by the Szeg\"o strong limit theorem, $\lim_{n \to \infty} D_n (\lambda) =e^{\lambda}$.
So we have $\kappa_n^2 \to 1$, as $n \to \infty$.

   From the orthonormal property and the recursion formula of the $\phi_n$'s, 
we show  that $\kappa_n$ satisfies the following difference-differential equation
\begin{equation}
 {n+1 \over 2 s} { (\kappa_n^2)_s \over \kappa_n^2}
 - { {\kappa_{n-1}^2 - \kappa_n^2} \over \kappa_n^2 }
 +{1\over4} { (\kappa_{n+1}^2)_s \over  \kappa_{n+1}^2}
 {(\kappa_n^2)_s \over  \kappa_{n+1}^2 -  \kappa_n^2} 
 - {1 \over 4} \left( { (\kappa_n^2)_s \over \kappa_n^2 }\right)^2 = 0, \label{E:1.2}
\end{equation}
where $s = \sqrt{\lambda}$, and $(\kappa_n^2)_s = {d \over ds} (\kappa_n^2).$
If we make the scaling
\begin{equation}
 \kappa_n^2 = 1 + {c_1 \over (n+1)^\alpha } R(T(n,s)),
\end{equation}
with $T(n,s) = t(n,s) + \epsilon(n, s)$, and
\begin{equation}
  t(n,s) = { (n+1)^\beta \over c_2} \left( {c_3 s \over n+1} -1 \right),
\end{equation}
then we obtain,  as $n \to \infty$, that the parameters satisfy
$\alpha=1/3, \beta = 2/3, c_1= - 2^{1/3}, c_2 = -1/2^{1/3}, c_3 = 2$, 
$\epsilon(n,s) =O(1/ (n+1)^{1/3})$, and the 
function $R$ satisfies
\begin{equation}
  (R^{''})^2 -8 (R')^3 +4 t (R')^2 -2 R' R^{'''} +o(1) =0. \label{E:1.6}
\end{equation}
If we drop the $o(1)$ term, the equation above is 
another form of the Painlev\'e II equation 
 which is discussed by Tracy and Widom in \cite{tracy}. 
So (\ref{E:1.2}) is called spatially discrete Painlev\'e II equation  in 
this sense.
   
   Therefore we have another proof of the asymptotic formula
\begin{equation}
  \kappa_n^2 = 1 - {2^{1/3} \over (n+1)^{1\over 3} } R(t)+\cdots, \label{E:1.10a}
\end{equation}
where $R'(t)=- q^2 (t)$, and $q(t)$ satisfies the Painlev\'e II equation
$q^{''}= t q + 2 q^3$. This asymptotic formula was first proved
by Baik, Deift and Johansson \cite{baik} using  Riemann-Hilbert problems.
They discussed much more asymptotic properties for
$\kappa_n$. See \cite{baik} for the details.
 The asymptotic formula (\ref{E:1.10a}) is  used for investigating the
distribution of the length  $l_n$ of the longest increasing subsequence
of a random permutation  \cite{baik}, 
by discussing the asymptotics of the 
$D_{n-1} (\lambda)= e^{\lambda} \prod_{k=n}^{\infty} \kappa_k^2$.
The distribution of $l_n$ (as $n \to \infty$) is same as 
the distribution of the largest eigenvalue of the
 Gaussian Unitary Ensemble (GUE) in
random matrix theory \cite{baik} \cite{tracy2} \cite{tracy3}. 
In \cite{tracy3}, Tracy and Widom used a different method to study 
the distribution of the $l_n$ by directly investigating the asymptotics
of $D_{n} (\lambda)$. They also obtained the distribution of
the length of the longest increasing subsequence of an odd permutatiom
\cite{tracy3}.

In \cite{hisakado} , Hisakado discussed the same polynomials 
as in this paper, and he also get  Painlev\'e II equation from
a discrete equation which is called discrete string equation in \cite{hisakado}. 
The difference is as follows.  In \cite{hisakado}, Hisakado discussed 
that  the constant term of polynomial
$\phi_n(z)/\kappa_n$ 
satisfies
a discrete equation (discrete string equation). 
In this paper, 
we discuss $\kappa_n$, the coefficient of the leading 
term of $\phi_n(z)$, which satisfies a spatially discrete equation 
(difference-differential equation). 
And the discrete string equation in \cite{hisakado} is  convergent to the 
original Painlev\'e II ( $q'' = x q + 2 q^3$). In this paper, the 
spatially discrete equation is   convergent to  a third order differential 
equation. As discussed by Tracy and Widom in \cite{tracy}, 
this third order equation 
 is equivalent to the Painlev\'e II equation. 

   It is known that the Painlev\'e equations have discrete analogs.
 See 
\cite{fokas} \cite{ramani} \cite{clarkson} \cite{jimbo} and references therein
for the discrete or q-difference Painlev\'e equations.
In this paper, we show that the Painlev\'e II equation has 
spatially discrte version. So far, to my knowlege,
the spatially discrete versions for Painlev\'e equations
are not known very well.
In \cite{fokas}, 
Fokas, Its and Kitaev discussed another spatially discrete integrable
equation, which is obtained from orthonormal polynomials on real line
satisfying a recursion formula in the form (\ref{E:2.40a}).
It is discussed in \cite{fokas} that the simultaneous solution of
the spatially discrete equation in \cite{fokas} 
and of discrete Painlev\'e I equation solves a special case of 
Painlev\'e IV equation. 

  This report is organized as follows. In the next section, we state
the recursion formula for the orthonormal polynomials on the unit circle,
which is proved in \cite{szego}.
In Sect.~3, we use the recursion formula and the orthonormal property
of the polynomials to investigate the leading coefficient $\kappa_n(s)$,
which is found to satisfy a difference-differential equation,
or spatially discrete equation. And we also get a formula for the
second leading coefficient of $\phi_n(z)$ in terms of $\kappa_n$.
In Sect.~4, the difference-differential equation is proved to tend to
the second Painlev\'e equation in a third order equation  form. 
Then we get the asymptotics of the coefficients for the leading 
term and for the second leading term of $\phi_n(z)$.

\setcounter{equation}{0}
\section{Recursion Formula for the Polynomials }

 Let us consider the orthonormal polynomials $\phi_n (n=0, 1,2, \cdots)$
on the unit circle in the complex plane  defined by (\ref{E:1.1}).
If we let $f(e^{i\theta})=\exp(2 \sqrt{\lambda} \cos \theta)$
 $= \exp(\sqrt{\lambda}(z+1/z))$, 
 $z=e^{i \theta}$, then (\ref{E:1.1}) becomes
\begin{equation}
 \int_{-\pi}^{\pi} \phi_n (e^{i \theta}) \overline{\phi_m (e^{i \theta})} 
 f(e^{i \theta}) {d\theta \over 2 \pi}
= \delta_{nm}, \,\,\,\,\,\, m,n \ge 0.  \label{E:2.1}
\end{equation}   
 Since  $2 \sqrt{\lambda} \cos \theta$ is even in $\theta$,
the coefficients of $\phi_n$ are real \cite{szego}.
It is also proved in \cite{szego} that $\phi_n$ satisfy the 
recursion formulas
\begin{eqnarray}
\kappa_{n-1} \,\,z \,\,\phi_{n-1}(z) &=& 
 \kappa_{n} \,\,\phi_{n}(z)-\phi_{n}(0) \,\,\phi_{n}^*(z),  \label{E:2.2} \\
\kappa_n \,\, \phi_{n+1}(z) &=& 
 \kappa_{n+1} \,\,z \,\,\phi_{n}(z)+\phi_{n+1}(0) \,\,\phi_{n}^*(z). \label{E:2.3}
\end{eqnarray}
Here the ``*'' is defined as
\begin{equation}
 \phi^*(z) = \overline{a_n} + \overline{a_{n-1}} \,\,z + \cdots+ \overline{a_0} \,\,z^n,  \label{E:2.4a}
\end{equation}
if $\phi(z) = a_n \,\,z^n+ a_{n-1} \,\,z^{n-1} + \cdots+a_0 $. Notice that
for $\phi_n (z)$, the coefficients are real as mentioned above.

   Since we are interested in the leading coefficient $\kappa_n$, we set 
$\phi_n(z)= \kappa_n\,p_n(z)$, and for simplicity we denote $s= \sqrt{\lambda}$. Then 
(\ref{E:2.1}) becomes
\begin{equation}
 \int_{-\pi}^{\pi} p_n (e^{i \theta}) \overline{p_m (e^{i \theta})} 
 f(e^{i \theta}) {d\theta \over 2 \pi}
= \delta_{nm} {1 \over \kappa_n^2} , \,\,\,\,\,\, m,n \ge 0.  \label{E:2.5}
\end{equation}   
If we eliminate $\phi_{n}^*(z)$ in (\ref{E:2.2}) and (\ref{E:2.3}), we get
\begin{equation}
z \left( {\kappa_{n+1} \over \kappa_n} \,\,\phi_n(z) -{\phi_{n+1}(0) \over \phi_n(0)}\,\,
 {\kappa_{n-1} \over \kappa_n} \,\,\phi_{n-1}(z) \right) =
 \phi_{n+1}(z) - {\phi_{n+1}(0) \over \phi_n(0)} \,\,\phi_{n}(z). \label{E:2.10}
\end{equation}
By $\phi_n(z)= \kappa_n\,\,p_n(z)$, (\ref{E:2.10}) becomes
\begin{equation}
z \left(  p_n(z) -{p_{n+1}(0) \over p_n(0)}
 {\kappa_{n-1}^2 \over \kappa_n^2} p_{n-1}(z) \right) =
 p_{n+1}(z) - {p_{n+1}(0) \over p_n(0)} p_{n}(z). \label{E:2.11}
\end{equation}
For the usage in later discussions, let us record this 
as a lemma, which is due to Szeg\"o \cite{szego}.
\begin{lemma}
 The orthogonal polynomials $p_n (z) $ defined by (\ref{E:2.5})
on the unit circle in the complex plane
satisfy the following recursion formula
\begin{equation}
z \,\,\Big( p_n(z) +C_n \,\,p_{n-1} (z) \Big) =p_{n+1}(z)+ B_{n} \,\,p_n(z), 
                     \label{E:2.20}
\end{equation}
where
\begin{eqnarray}
 C_n &=& - {p_{n+1}(0) \over p_n(0)} \,\,{\kappa_{n-1}^2 \over \kappa_n^2}, \label{E:2.25} \\
 B_n &=& - {p_{n+1}(0) \over p_n(0)}.  \label{E:2.26}
\end{eqnarray}
\end{lemma}
  As a remark, the recursion formula for the orthogonal polynomials on the real line
  takes the form \cite{szego} \cite{tracy1}
\begin{equation}
 z \,\,p_n = a_n \,\,p_{n+1} + b_n \,\,p_{n}+c_n\,\, p_{n-1}. \label{E:2.40a}
\end{equation}
But we have seen that on the unit circle with the weight $d \mu$, 
 the coefficient of $p_{n-1}$ 
in the recursion
formula contains $z$.
\setcounter{equation}{0}
\section{Spatially Discrete Equation for $\kappa_n$ }

In this section, we use the orthogonal
relation (\ref{E:2.5}) and the recursion relation (\ref{E:2.20}) to show that
$\kappa_n(s)$ satisfies a  spatially
discrete equation (difference-differential equation).
In the next section we show that
the contineous limit of this spatially discrete equation gives a third order 
ordinary differential equation, which is equivalent to 
Painlev\'e II equation. 

  Let us  write (\ref{E:2.5}) in the form
\begin{equation}
 <p_n, p_m>
\equiv  \oint p_n (z) \overline{p_m (z)}  f(z) {dz \over z}
= \delta_{nm}\, h_n, \,\,\,\,\,\, m,n \ge 0 , \label{E:3.5}
\end{equation}
where $h_n = 2 \pi i /\kappa_n^2$, and the integral $\oint$ is 
taken over the unit circle in the complex plane. We will use the notation
$d \mu = f(z) dz/ z$ for simplicity.
First it is  easy to show the following identities
\begin{eqnarray}
 \oint z^k \,{\partial \overline{p_m(z)} \over \partial \bar{z}} \, p_n(z) d\mu &=&
 \oint z^{-k} \,{\partial p_m(z) \over \partial z} \, \overline{p_n(z)} d\mu,  \label{E:3.10} \\
 \oint z^k \, \overline{p_m(z)}  \, p_n(z) d\mu &=&
 \oint z^{-k} \, p_m(z)  \, \overline{p_n(z)} d\mu,  \label{E:3.11}
\end{eqnarray}
by using $z=e^{i \theta}$, and $\theta$ replaced by $-\theta$. Recall 
$\overline{p_n(z)} = p_n(\bar{z})$.

   Now, consider (\ref{E:3.5}) with $m=n$,
\begin{eqnarray*}
 h_n &=& \oint p_n (e^{i \theta}) \bar{p_n (e^{i \theta})} 
  e^{ s (z + {1\over z}) } {dz \over z} \\
  &=& {1 \over s} \oint p_n (e^{i \theta}) \,\bar{p_n (e^{i \theta})} \, 
  e^{ s  {1\over z} }  \,{1 \over z} \,d (e^{s z}).
\end{eqnarray*}
Then by integration by parts and using (\ref{E:3.10}) and (\ref{E:3.11}),
we have
\begin{equation}
 h_n = {1 \over s} \oint z^2 \, {\partial p_n(z) \over \partial z} \, \overline{p_n(z)} d\mu + 
    \oint  z^2 p_n(z) \overline{p_n(z)} d \mu + 
  {1 \over s} \oint  z p_n(z) \overline{p_n(z)} d \mu .  \label{E:3.15}
\end{equation}

  Since $\overline{p_n(z)}=p(\bar{z})=p(1/z)$ on the unit circle, by the Cauchy
integral formula, the right hand side of the above equation can be
calculated, and the result involves all the coefficients of $p_n(z)$. 
That does not help too much to investigate the $\kappa_n$. Here, by using
the recursion formula and the orthonormal property, the right side of
(\ref{E:3.15}) can be expressed in terms of $h_{n-1}, h_n, h_{n+1}$ and
their first  derivatives. 

  We need to consider two type integrals, 
$$\oint z^k\,{\partial \over \partial z} p_n(z)\, 
 \overline{p_n(z)}\,d \mu, \,\,\,\,\,\oint z^k\,p_n(z)\, \overline{p_n(z)}\,d \mu$$ 
for (\ref{E:3.15}), where $k$ is integer. In this paper, we do not discuss the 
general formula. 
We just consider  the integrals in (\ref{E:3.15}), and the formulas are
given in the following lemmas.
\begin{lemma}
\begin{equation}
\oint  z\, p_n(z)\,\overline{p_n(z)}\, d\mu=(B_n -C_n) \,h_n.
\end{equation}
\end{lemma}
{\it Proof.} Multiply $\overline{p_n(z)}$ on both sides of (\ref{E:2.20}).
Then  integrating on both sides yields the result.
\rule{1.6ex}{1.6ex}
\begin{lemma}
\begin{equation}
 \oint z^2 p_n(z) \overline{p_n(z)} d \mu  =
-(C_{n+1}\,(B_n-C_n)\,-(B_n-C_n)^2\,+C_n\,(B_{n-1}-C_{n-1}))\,h_n.
\end{equation}
\end{lemma}
{\it Proof.}
Let $I = \oint z^2 \,p_n(z)\,\overline{p_n(z)} d\mu.$  By Lemma 1, we have
$$ z^2 \,p_n = p_{n+2}+B_{n+1} \,p_{n+1}-C_{n+1}\,z\,p_n+B_n\,z\,p_n-C_n\,z^2\,p_{n-1},$$
which implies 
\begin{equation}
I=\oint \left( (-C_{n+1} +B_n)\,z\,p_n-C_n\,z^2\,p_{n-1} \right)\,\overline{p_n(z)}\,d\mu,
   \label{E:3.7a}
\end{equation}
where we have used $<p_{n+2},p_n>=<p_{n+1},p_n>=0$. 
Again by Lemma 1, 
$$
z^2\,p_{n-1} = z\,p_n+B_{n-1}\,p_n+B_{n-1}^2\,p_{n-1}-z\,B_{n-1}\,C_{n-1}\,p_{n-2}
-C_{n-1}\,z^2\,p_{n-2}.$$
Since  $<p_{n-1}, p_n>=<z\,p_{n-2}, p_n>=0$, we then obtain from (\ref{E:3.7a})
\begin{eqnarray*}
I&=&\oint \left \{ (-C_{n+1}+B_n)\,z\,p_n-C_n\,(z\,p_n+B_{n-1}\,p_n-C_{n-1}\,z^2\,p_{n-2})
      \right \} \,\overline{p_n} d\mu \\
 &=& \oint {\Big \{ }  (-C_{n+1}+B_n-C_n)\,(p_{n+1}+B_n\,p_n-z\,C_n\,p_{n-1})   \\
 &&\hspace{4cm}   -C_n\,B_{n-1}\,p_n+C_n\,C_{n-1}\,z^2\,p_{n-2}
    {\Big \} }  \,\overline{p_n} d\mu \\
 &=&  (-C_{n+1}\,+B_n-C_n)(B_n-C_n)\,h_n - C_n\,B_{n-1}\,h_n+C_n\,C_{n-1}\,h_n, 
\end{eqnarray*}
where $<z\,p_{n-1}, p_n>=<z^2\,p_{n-2}, p_n>=h_n$.
\rule{1.6ex}{1.6ex}\\

  In order to evaluate the integral $<z^2 \, \partial p_n/\partial z, p_n>$ in (\ref{E:3.15}), 
we need to consider the 
second leading coefficient of $p_n(z)$. Suppose
\begin{equation}
  p_n (z) = z^n + b_{n-1}\, z^{n-1} + \cdots.  \label{E:3.30}
\end{equation}
Because $ z^2\, {\partial p_n (z) \over \partial z} =n\,z^{n+1}+(n-1)\,b_{n-1}\,z^{n}+\cdots,$
we set 
$$ z^2\, {\partial p_n (z) \over \partial z} =n(p_{n+1} + \mu_n\, p_n +\cdots).$$ 
Then by
\begin{eqnarray*}
 z^{n+1}&+&b_n \, z^n +\cdots \\
   &+& \mu_n (z^n +b_{n-1} \,z^{n-1}+\cdots) \\
 = z^{n+1} &+&(1-{1\over n}) \,b_{n-1}\,z^n+\cdots,
\end{eqnarray*}
it follows that
\begin{equation}
 \mu_n =\left(1-{1\over n} \right) b_{n-1}-b_n.
\end{equation}
Since $<z^2\, {\partial p_n \over \partial z},  p_n>=n\,\mu_n \,h_n$, 
we have proved the following lemma.
\begin{lemma}
\begin{equation}
\oint z^2 \, {\partial p_n(z) \over \partial z} \, \overline{p_n(z)} d\mu =
\left( n\,(b_{n-1}-b_n) - b_{n-1} \right) h_n.
\end{equation}
\end{lemma}

   Because we want  an equation for the leading coefficient $\kappa_n$,
we need to express the second leading coefficient $b_n$ 
in terms of $\kappa_n$. First, the $b_n$ can be expressed in terms of
$B_n$ and $C_n$.
\begin{lemma} For the second leading coefficient $b_n$ of $p_n$ defined by
(\ref{E:3.30}), there are the following properties
\begin{eqnarray}
 \hspace{-3cm} &(i) \hspace{2cm} & b_{n-1}-b_n =B_n - C_n, \label{E:3.40}  \\
 \hspace{-3cm} &(ii) \hspace{2cm} &{1 \over s}\,b_{n-1}=
   - C_{n}\,(B_{n-1}-C_{n-1})-{h_{n} \over h_{n-1}}.  \label{E:3.41}
\end{eqnarray}
\end{lemma}
{\it Proof.}
By Lemma 1, $z \,\,p_n(z)  =p_{n+1}(z)+ B_{n} \,\,p_n(z)-C_n \,z\,\,p_{n-1} (z)$.
By (\ref{E:3.30}), $z\, p_n=p_{n+1}+(b_{n-1}-b_n) \,p_n+\cdots$. Then we
have $(b_{n-1}-b_n)h_n =B_n\,h_n - C_n\,h_n$ by considering $<z\,p_n, p_n>$ in the two ways.

  To prove the second formula (\ref{E:3.41}), consider
\begin{equation}
I=\oint p_{n+1}(z)\,\overline{p_n(z)} \,\,e^{s (z+{1\over z})}\,dz. 
\end{equation}
Using integration by parts, 
\begin{eqnarray*}
I &=& {1\over s} \oint p_{n+1}(z)\,\overline{p_n(z)} \,\,e^{s {1\over z}}\,d e^{s z} \\
 &=&{-1 \over s} \oint (z\, {\partial p_{n+1} \over \partial z})\,\overline{p_n}\,d\mu
  + {1\over s} \oint p_{n+1}\,\left(\bar{z}\,
  {\partial \overline{p_n} \over \partial\bar{z}} \right) d\mu
+\oint p_{n+1} \,(\bar{z}\,\overline{p_n} )\,d\mu,
\end{eqnarray*} 
where $<p_{n+1}, z\,{\partial p_n \over \partial z}>=0$, and
$<p_{n+1}, z\,p_n>=h_{n+1}$. And 
$z\, {\partial p_{n+1} \over \partial z}=(n+1)(z^{n+1}+{n\over n+1}\,b_n\,z^n+\cdots)$.
Set $z\, {\partial p_{n+1} \over \partial z}=(n+1)(p_{n+1}+ \gamma_n \, p_n +\cdots)$,
which implies 
$<z\, {\partial p_{n+1} \over \partial z}, p_n>= (n+1)\,\gamma_n\,h_n$ . 
Then by 
$$p_{n+1} + \gamma_n p_n +\cdots = z^{n+1} +{n \over n+1} b_n z^n +\cdots,$$
we get $\gamma_n + b_n ={n \over n+1} b_n$, so
$\gamma_n =-{1\over n+1} \, b_n$. Thus
\begin{equation}
I={1\over s} \, b_n\,h_n+h_{n+1}.  \label{E:3.50}
\end{equation}

  On the other hand, the integral $I$ can be evaluated by using the recursion formula 
(see (\ref{E:2.20}))
$$
z\, p_{n+1}= p_{n+2}+B_{n+1}\,p_{n+1}-C_{n+1}\,p_{n+1}-C_{n+1}\,B_n\,p_n+ C_{n+1}\,C_n\,z\,p_{n-1},
$$
which gives
\begin{equation}
I=<z\,p_{n+1}, p_n>= -C_{n+1}\,( B_n -C_n)\,h_n. \label{E:3.55}
\end{equation}
By (\ref{E:3.50}) and (\ref{E:3.55}), we get (\ref{E:3.41}).
\rule{1.6ex}{1.6ex} \\

  We have seen  by the last four lemmas that the right hand side of (\ref{E:3.15}) can
be expressed in terms of $B_n$ and $C_n$. The following lemma gives the
formulas of $B_n, C_n$ in terms of $\kappa_n$.
\begin{lemma}
 The coefficients $B_n, C_n$ in the recursion formula (see (\ref{E:2.25}),
(\ref{E:2.26})) can be expressed as follows.
\begin{eqnarray}
 \hspace{-4cm} &(i) \hspace{3cm} &B_n - C_n  = -{(\kappa_n^{2})_s \over 2 \,\kappa_n^2}, 
          \label{E:3.51} \\
 \hspace{-4cm} &(ii) \hspace{3cm} & B_n = - {(\kappa_n^2)_s \over 2\, (\kappa_n^2-\kappa_{n-1}^2)}, 
      \label{E:3.52} \\
 \hspace{-4cm} &(iii) \hspace{3cm} & C_n =-{\kappa_{n-1}^2 \over 2\, \kappa_n^2} \,
      { (\kappa_n^2)_s \over \kappa_n^2 - \kappa_{n-1}^2 },  \label{E:3.53}
\end{eqnarray}
where $\kappa_n$ is the leading coefficient of $\phi_n$ defined by (\ref{E:1.1}), and
$(\kappa_n^2)_s = {d \kappa_n^2(s) \over ds}$.
\end{lemma}
{\it Proof.}
Since $h_n=<p_n, p_n>$, we have 
\begin{equation}
{d h_n(s) \over ds} = \oint p_n(z)\,\overline{p_n(z)} \, \left(z+{1\over z}\right) d\mu=
  2 \oint z\,p_n\, \overline{p_n}\, d\mu,
\end{equation}
where we have used 
$<{\partial p_n \over \partial s}, p_n>=<p_n, {\partial p_n \over \partial s}>=0$, since
deg$({\partial  p_n \over \partial s}) \le n-1$,  
deg$({\partial \overline{p_n} \over \partial s}) \le n-1$.
Then by Lemma 1 and the notation $h_n=2 \pi i/\kappa_n^2$, we have the first formula (\ref{E:3.51}).
By (\ref{E:2.25}) and (\ref{E:2.26}), there is $C_n = B_n \, \kappa_{n-1}^2/ \kappa_n^2$.
We then get (\ref{E:3.52}) and (\ref{E:3.53}).
\rule{1.6ex}{1.6ex} \\

  By Lemma 2, 3, 4, (\ref{E:3.15}) is changed to
\begin{equation} \begin{array}{ll}
1=& \displaystyle{n \over s} ( b_{n-1}-b_n ) -\displaystyle{1\over s}b_{n-1} \\
 &-C_{n+1}(B_n-C_n)+(B_n-C_n)^2-C_n(B_{n-1}-C_{n-1}) +\displaystyle{1\over s} (B_n-C_n). \end{array}
\end{equation}
By Lemma 5, this equation  further becomes
\begin{equation}
{n+1 \over s} ( B_n -C_n ) +{h_n \over h_{n-1} }-1 
 -C_{n+1}(B_n-C_n)+(B_n-C_n)^2 =0.
\end{equation}
Therefore by Lemma 6, we have proved the following theorem.
\begin{theorem}
The leading coefficient $\kappa_n(s)$ of the orthonormal polynomials $\phi_n(z)$
defined by (\ref{E:1.1}) satisfies the following 
spatially discrete equation ( difference-differential equation)
\begin{equation}
 {n+1 \over 2 s} { (\kappa_n^2)_s \over \kappa_n^2}
 - { {\kappa_{n-1}^2 - \kappa_n^2} \over \kappa_n^2 }
 +{1\over4} { (\kappa_{n+1}^2)_s \over  \kappa_{n+1}^2}
 {(\kappa_n^2)_s \over  \kappa_{n+1}^2 -  \kappa_n^2} 
 - {1 \over 4} \left( { (\kappa_n^2)_s \over \kappa_n^2 }\right)^2 = 0, \label{E:3.f}
\end{equation}
where $s = \sqrt{\lambda}$, and $(\kappa_n^2)_s = {d \over ds} (\kappa_n^2).$
\end{theorem}
  This is a new result for $\kappa_n(s)$, the leading coefficients of 
the orthonomial polynomials $\phi_n(z, s)$, with the weight 
$\exp \Big( s (z +{1 \over z}) \Big) dz$ on the unit circle.
In the next section, we show that as  $n \to \infty$, this equation 
is reduced to a third order ordinary differential equation which is 
equivalent to the Painlev\'e II equation.
\setcounter{equation}{0}
\section{Painlev\'e II Equation }
   We have shown that $\kappa_n^2$ satisfies the equation (\ref{E:3.f}). As
mentioned in Sect.~1,  $\kappa_n^2$ satisfies the boundary condition
$\kappa_n^2 =1 +o(1)$, as $n \to \infty$. In this section, we  
compute the asymptotics of $\kappa_n^2 -1$, as $n \to \infty$. We will see that
the asymptotics involves the second Painlev\'e function.

  Equation (\ref{E:3.f}) has two independent variables $n$ and $s$. To study 
the asymptotics, we use ``similarity'' reduction, or the so
called double scaling method. By comparing
the first two terms in (\ref{E:3.f}), we consider the case when $n+1$ and $s$ are in
the same order as they are large, and let
\begin{eqnarray}
 {c_3 \,s \over n+1 } &=& 1 + {c_2 t \over (n+1)^{\beta}},  \label{E:4.1} \\
 \kappa_n^2 &=&1+ {c_1 \over (n+1)^{\alpha} } R(T(n, s)),  \label{E:4.2}
\end{eqnarray} 
where $T(n,s) =  t(n,s) + \epsilon(n, s)$,  the leading term $t(n,s)$ 
is defined by (\ref{E:4.1}),
and it will be shown that $\epsilon(n,s)$ is  a smaller term as $n \to \infty$.
We want to determine the constants $\alpha, \beta, c_1, c_2, c_3$,
such that as $n \to \infty$, (\ref{E:3.f}) is reduced to a differential equation.

Let us consider the approximate expressions of
 $(\kappa_n^2)_s$, $\kappa_{n+1}^2 - \kappa_n^2$ in
terms of $R'$  and maybe also higher order derivatives if we need.
Here $R'$ means the limit of $(R(T+\Delta T)-R(T))/\Delta T$.
By (\ref{E:4.1}), for fixed $s$, as $n$ increases by $1$, $t$ is increased by 
a order $1/(n+1)^{1-\beta}$.
For the asymptotics of $\kappa_{n+1}^2 -\kappa_n^2$ and  
$(\kappa_{n+1}^2)_s - (\kappa_n^2)_s$,
the higher order terms also need to be concerned in order to
have all the terms in the first three leading orders in the expansion of left side of
(\ref{E:3.f}).
Therefore we have the following
\begin{eqnarray*}
\hspace{-0.5cm}& & t(n+1, s) - t(n, s) = {c_4 \over (n+1)^{1-\beta} } + \cdots, \\
\hspace{-0.5cm}& & \kappa_{n+1}^2 -\kappa_n^2 = R^{'}\,{c_5 \over (n+1)^{1+\alpha-\beta} } 
 + R^{''}\, {c_6 \over (n+1)^{2+ \alpha- 2 \beta} } \\
\hspace{-0.5cm}&& \hspace{5.3cm} + R^{'''}\,{c_7 \over (n+1)^{3+\alpha-3 \beta} } 
 + R \, {c_8 \over (n+1)^{1 +\alpha} }  + \cdots, \\
\hspace{-0.5cm}& & \kappa_{n-1}^2 -\kappa_n^2 = - R^{'} \,{c_5 \over (n+1)^{1+\alpha-\beta} } 
 + R^{''} \,{c_6 \over (n+1)^{2+ \alpha- 2 \beta} } \\
\hspace{-0.5cm}&& \hspace{5.7cm} - R^{'''} \, {c_7 \over (n+1)^{3+\alpha-3 \beta} } 
 - R \,{c_8 \over (n+1)^{1 +\alpha} }  + \cdots, \\
\hspace{-0.5cm}&& (\kappa_n^2 )_s = {c_1 c_3 \over c_2} \, R' \,  
  {1 \over (n+1)^{1 +\alpha- \beta} } + \cdots, \\
\hspace{-0.5cm}& & (\kappa_{n+1}^2)_s - (\kappa_n^2)_s = { c_1 c_3 \over c_2 (n+1)^{1+\alpha- \beta} }
 \left(  R^{''}\, {c_4 \over (n+1)^{1-\beta} }
 +{R^{'''} \over 2} \, {c_4^2 \over (n+1)^{2-2 \beta} } \right) +\cdots,
\end{eqnarray*}
where $c_4$ is a constant, and
\begin{eqnarray}
   & c_5 =c_1\, c_4,  & c_6={1\over2}\,c_1 \,c_4^2,  \label{E:4.30} \\
   & c_7=\displaystyle{1\over6} \,c_1 \,c_4, & c_8 = - c_1\, \alpha.  \label{E:4.31}
\end{eqnarray}

  Now  write (\ref{E:3.f}) in the following form  
$$ \begin{array}{ll}
 &\displaystyle{n+1 \over 2 s} (\kappa_n^2)_s (\kappa_{n+1}^2 -\kappa_n^2) 
 -(\kappa_{n-1}^2-\kappa_n^2)(\kappa_{n+1}^2 - \kappa_n^2)
+{1\over 4}(\kappa_n^2)_s (\kappa_{n+1}^2)_s \\ \vspace{0.3cm}
& \hspace{3cm} - \displaystyle{1\over 4} \displaystyle{\kappa_{n+1}^2-\kappa_n^2 \over \kappa_{n+1}^2} 
   (\kappa_n^2)_s (\kappa_{n+1}^2)_s
 - \displaystyle{1\over 4} \displaystyle{ ( (\kappa_n^2)_s )^2  \over \kappa_n^2 } 
 (\kappa_{n+1}^2 -\kappa_n^2) =0.
\end{array} $$
 By  substituting the asymptotic formulas above into this equation,   we get
\begin{equation}
  S_1 +S_2 + S_3 + S_4 +o(1)=0,  \label{E:4.40}
\end{equation}
where
\begin{eqnarray*}
 S_1& =&{c_1 c_3^2 \over 2 c_2} \left( 1-{c_2 t \over (n+1)^\beta} \right) 
    { R^{'}\over (n+1)^{1+\alpha-\beta}}  \times \\
 && \hspace{0.3cm} \left \{{c_5 R^{'} \over (n+1)^{1+\alpha-\beta}} 
   + {c_6 R^{''} \over (n+1)^{2+\alpha-2 \beta} } 
   + {c_7 R^{'''} \over (n+1)^{3+\alpha-3 \beta}} 
   + {c_8 R \over (n+1)^{1+\alpha} }  \right \},\\ \vspace{0.3cm}
 S_2& =&  {c_5^2 (R^{'})^2 \over (n+1)^{2 +2 \alpha -2\beta} }
    + {2 c_5 c_7 R^{'} R^{'''} \over (n+1)^{4+2 \alpha -4 \beta} } 
      + { 2 c_5 c_8 R^{'} R \over (n+1)^{2+2 \alpha -\beta}  } 
    - { c_6^2 (R^{''})^2 \over (n+1)^{4+2 \alpha -4 \beta}  } ,  \\ \vspace{0.3cm}
 S_3& =& {c_1 c_3 \over 4 c_2}  {R^{'} \over (n+1)^{1+\alpha-\beta} } \times \\
  && \left \{ {c_1 c_3 \over c_2}{R^{'} \over (n+1)^{1+\alpha-\beta} } 
  + {c_1 c_3 \over c_2 (n+1)^{1+\alpha-\beta} }
 \left( {c_4 R^{''} \over (n+1)^{1-\beta} } +{c_4^2 R^{'''} \over 2 (n+1)^{2-2 \beta} }
        \right) \right \},\\ \vspace{0.3cm}
  S_4& =& -{c_5\over4 \kappa_{n+1}^2} \left( {c_1 c_3 \over c_2} \right)^2 
  { (R^{'})^3 \over  (n+1)^{3+3 \alpha -3 \beta} }
  - {c_5\over4 \kappa_{n}^2} \left( {c_1 c_3 \over c_2} \right)^2 
   {(R^{'})^3 \over (n+1)^{3+3 \alpha -3 \beta} }.
\end{eqnarray*}
By the definition of $t$ and asymptotics of $\Delta t$, we have $0< \beta < 1$.
Since $\kappa_n^2 \to 1$, as $n \to \infty$, we have $\alpha >0$.
Consider the orders of the terms on the left side of (\ref{E:4.40}).
The coefficients of $(R^{'})^2, R^{'} R^{''}, R^{'} R^{'''}, (R^{''})^2$ have orders
$2(1+\alpha-\beta)$, $2(1+\alpha-\beta)+(1-\beta)$, 
$ 2(1+\alpha-\beta)+2(1-\beta)$,  $2(1+\alpha-\beta)+2(1-\beta)$ respectively.
The coefficients of $t (R^{'})^2, R R^{'}, (R^{'})^3$ have  orders
$2(1+\alpha-\beta)+\beta$, $2(1+\alpha-\beta)+\beta$,  
$2(1+\alpha-\beta)+(1+ \alpha - \beta)$ respectively. 
And the $o(1)$ in (\ref{E:4.40})
contains  higher order terms which we do not concern.
To determine the values of $\alpha$ and $\beta$, the only choice is to set
 the coefficients of $R^{'} R^{'''}$, $(R^{''})^2$,  $t (R^{'})^2$, $R R^{'}$ and $(R^{'})^3$
 be in the same order.
So we have  
$2(1-\beta) = \beta = 1+\alpha - \beta$.
The solution is unique $ \alpha =1/3, \beta = 2/3$. So (\ref{E:4.40}) becomes
\begin{equation}
\begin{array}{lll}
 && A_1 \,\displaystyle{(R^{'})^2 \over (n+1)^{4/3} } 
 + A_2  \,\displaystyle{R^{'} R^{''} \over (n+1)^{5/3} } \\
&& + \left( A_3 \,R^{'} R + A_4 \,R^{'} R^{'''}  + A_5 \,(R^{''})^2 +
 A_6 \, t (R^{'})^2 + A_7 \,(R^{'})^3 \right)  \displaystyle{1\over (n+1)^2 } \\
&& \hspace{2cm} + O\left(  \displaystyle{ 1\over (n+1)^{7/3} } \right) =0, 
\end{array} \label{E:4.50}
\end{equation}
where 
\begin{eqnarray*}
A_1 &=& {c_3 \over 2 } \,{c_1 c_3 \over c_2} \,c_5 +c_5^2 
         +{1 \over 4} \,\left({c_1 c_3 \over c_2} \right)^2, \\
A_2 &=& {c_3 \over 2} \,{c_1 c_3 \over c_2 } \,c_6 
        +{1\over4} \,\left({c_1 c_3 \over c_2} \right)^2 \,c_4,  \\
A_3 &=&  {c_3 \over 2} {c_1 c_3 \over c_2} c_8 + 2 c_5 c_8, \\
A_4 &=& {c_3 \over 2} \,{c_1 c_3 \over c_2} \,c_7 + 2 \,c_5 \,c_7 
 +{1\over 8} \,\left({c_1 c_3 \over c_2} \right)^2 \,c_4^2, \\
A_5 &=& -c_6^2, \\
A_6 &=& -{1\over 2} c_1 c_3^2 c_5, \\
A_7 &=& -{1\over 2} \,\left({ c_1 c_3 \over c_2 }\right)^2 \,c_5.
\end{eqnarray*}

   Look at the equation (\ref{E:4.50}). If $A_1$ or $A_2$ is not zero, 
then we get asymptotics of $\kappa_n$ contradicting to the 
results proved in \cite{szego}. In \cite{szego}, 
the asymptotics of $\kappa_n$ contains  exponential
function of $s$. 
So we must have $A_1 = A_2 =0$. 
The coefficient of the $1/(n+1)^2$ in ({\ref{E:4.50}) has the
pattern of Painlev\'e II equation  in \cite{tracy}.
So we want to choose proper constant numbers 
$c_1, c_2, c_3, c_4$, such that as $n \to \infty$, ({\ref{E:4.50})
reduces to  the Painlev\'e II equation.
In \cite{tracy}, Tracy and Widom  discussed two forms of 
 Painlev\'e II equation
\begin{equation}
{1\over2} {R^{'''} \over R^{'}} -{1\over2} {(R^{''})^2 \over (R^{'})^2 }
- { R \over R^{'} } +R^{'} =0,  \label{E:4.60}
\end{equation}
\begin{equation}
(R^{''})^2 + 4 R^{'} ( (R^{'})^2 - t R^{'} + R ) =0, \label{E:4.61}
\end{equation}
where $R^{'}(t) = - q(t)^2$, and $q(t)$ satisfies the original
Painlev\'e II $q^{''} = t q + 2 q^3$.
Equation (\ref{E:4.61}) is called Jimbo-Miwa-Okamoto $\sigma$ form
for Painlev\'e II.
Eliminating the $R$ in (\ref{E:4.60}) and (\ref{E:4.61})
gives another form
\begin{equation}
 - 2 \,R^{'} R^{'''}+ (R^{''})^2  + 4 \,t \,(R^{'})^2  - 8 (R^{'} )^3 =0.
       \label{E:4.62}
\end{equation}
 In (\ref{E:4.50}), we set $A_1=A_2=A_3=0$, $A_4=-2 A_5$, 
$A_6 = 4 A_5$, and  $A_7 = -8 A_5$.
By using (\ref{E:4.30}) and (\ref{E:4.31}), the solution of 
these nonlinear equations is 
found to be
\begin{eqnarray}
 &&c_1 = - 2^{1/3}, \\
 &&c_2 = - {1 \over 2^{1/3} },  \\
 &&c_3^2 = 4, \\
 && c_4 = 2^{1/3}.
\end{eqnarray}
Since we consider positive $n$ and $s$, $c_3$ is positive, i.e., $c_3 = 2$.  
And it is seen that 
$\epsilon(n, s) = O(1/(n+1)^{1/3})$ because the last term 
in (\ref{E:4.50}) is $1/3$ order higher than the 
proceding term. 

  Therefore we have a formal proof of the following theorem which was first
proved by Baik, Deift and Johansson \cite{baik} by using Riemann-Hilbert method.
\begin{theorem}
As $n \to \infty$, $\kappa_n^2$ has the following asymptotic formula
\begin{equation}
  \kappa_n^2 = 1 - {2^{1/3} \over (n+1)^{1/ 3} } R(t)
+ O \left( {1 \over (n+1)^{2/3} } \right),
\end{equation}
where $t$ is defined by 
${2 \sqrt{\lambda} \over n+1 }= 1 - {t \over 2^{1/3} (n+1)^{2/3} }$,
 $R'(t) = -q^2(t)$, and $q(t)$ satisfies Painlev\'e II $q^{''}=t q +2 q^3$.
\end{theorem}

 As discussed in \cite{baik}, the Painlev\'e II function $q(t)$ in Theorem 2
satisfies the boundary condition $q(t) \sim - Ai(t)$, as $t \to \infty$, where
$Ai(t)$ is the Airy function. 
This boundary condition can also be obtained by the asymptotics of
$\kappa_n$ in terms of exponential function \cite{szego} and the
Painlev\'e II equation that $q(t)$ satisfies.
The Painlev\'e II function $q(t)$ with
this boundary condition is discussed by Hastings and McLeod in \cite{mcleod}.

 As a remark, there is no resolution 
if ones want to get (\ref{E:4.60}) or (\ref{E:4.61}) from (\ref{E:4.50}).
So these three equations (\ref{E:4.60}), (\ref{E:4.61}) and (\ref{E:4.62})
play different roles. It is interesting that in \cite{haine} there is 
a similar augument for three forms of Painlev\'e VI 
equation.
It is discussed in \cite{haine} that by eliminiting the third order derivative
in  the equation obtained in \cite{haine} and in the equation in \cite{tracy1},
ones get the $\sigma$ form of Painlev\'e VI. 

Finally by Lemma 5, Lemma 6 and 
Theorem 2, we obtain the asymptotics for the second leading coefficient of
$\phi_n(z)$.
\begin{theorem}
For the polynomial $\phi_n(z)= \kappa_n \,(z^n + b_{n-1} \, z^{n-1} +\cdots )$
defined by (\ref{E:1.1}), the second leading coefficient 
$\kappa_n\, b_{n-1}$ has the asymptotic formula
\begin{equation}
{\kappa_n \, b_{n-1} \over \sqrt{\lambda} } 
= -1 + {1 \over 2^{2/3} (n+1)^{1/ 3} }  R(t) 
+ O \left( {1 \over (n+1)^{2/3} } \right),
\end{equation}
as $n \to \infty$, where $t$ and $R(t)$ are same as in Theorem 2.
\end{theorem}
\section*{Acknowledgements}
The author sincerely thanks Professor Craig A. Tracy 
for introducing this problem and 
the relevant reference papers,
 and  especially for helpful discussions for the calculations in Sect. 4. 
We also thank Professor Joel Keizer, Director of ITD, for providing 
very good working facilities.


\begin{thebibliography}{99}
  
  \bibitem{szego}
   G. Szeg\"o, {\it Orthogonal Polynomials}, American Mathematical Society
   Colloquium Publications, Vol 23, 4th Ed, New York, 1975. 

  \bibitem{tracy}
  C. A. Tracy and H. Widom, {\it Level-spacing distributions and the Airy
  kernel}, Comm. Math. Phys., {\bf 159} (1994), $151-174$.

  \bibitem{baik}
   J. Baik, P. Deift and K. Johansson, {\it On the distribution of the length
    of the longest increasing subsequence of random permutations}, preprint,
   LANL archives, math. CO/9810105.  
 
  \bibitem{tracy2}
  C. A. Tracy and H. Widom, {\it Universality of the distribution functions 
   of random matrix theory}, preprint.

  \bibitem{tracy3}
  C. A. Tracy and H. Widom, {\it Random unitary matrices, permutations 
  and Painlev\'e }, preprint.

  \bibitem{hisakado}
   M. Hisakado, {\it Unitary matrix model and Painlev\'e III}, 
   Mod. Phys. Lett. {\bf A11} (1996), $3001-3010$.

  \bibitem{fokas}
   A. S. Fokas, A. R. Its and A. V. Kitaev, {\it Discrete Painlev\'e equations
    and their appearance in quantum gravity,} Comm. Math. Phys. {\bf 142} (1991), 
   $313-344$.

  \bibitem{ramani}
   A. Ramani, B. Grammaticos and J. Hietarinta,
   {\it Discrete versions of the Painlev\'e equations,} 
   Phys. Rev. Lett. {\bf 67} (1991), $1829-1832$.

  \bibitem{clarkson}
   A. Bassom and A. P. Clarkson, {\it New exact solutions of the discrete fourth
   Painlev\'e equation.} Phys. Lett. A, {\bf 194} (1994), no.5-6, $358-370$.

  \bibitem{jimbo}
   M. Jimbo and H. Sakai, {\it A q-analog of the sixth Painlev\'e equation,}
   Lett. Math. Phys. {\bf 38} (1996), $145-154$.

  \bibitem{tracy1}
   C. A. Tracy and H. Widom, {\it Fredholm determinants, 
   differential equations  and matrix models}, 
   Comm. Math. Phys.,  {\bf 163} (1994), $33-72$. 

  \bibitem{mcleod}
   S. P. Hastings and J. B. McLeod, {\it A boundary value problem associated with
   the second Painlev\'e transcendent and the Korteweg de Vries equation},
   Arch. Rat. Mech. Anal. {\bf 73} (1980), $31-51$.

  \bibitem{haine}
   L. Haine and J.-P. Semengue,  {\it The Jacobi polynomial ensemble and
     the Painlev\'e VI equation,} preprint.

\end{thebibliography}
\end{document}